%% file: B2G-12-015_temp.tex
\pdfoutput=1

\documentclass[11pt,twoside,a4paper,cmspaper,final,collab]{cms-tdr}
\def\svnVersion{224468}\def\cmsCernNo{2013-215}\def\cmsCernDate{\today}\def\cmsMessage{Published in Physics Letters B as \href{http://dx.doi.org/10.1016/j.physletb.2014.01.006}{\doi{10.1016/j.physletb.2014.01.006.}}}

\begin{document}\cmsNoteHeader{B2G-12-015}

\hyphenation{had-ron-i-za-tion}
\hyphenation{cal-or-i-me-ter}
\hyphenation{de-vices}

\RCS$Revision: 222084 $
\RCS$HeadURL: svn+ssh://svn.cern.ch/reps/tdr2/papers/B2G-12-015/trunk/B2G-12-015.tex $
\RCS$Id: B2G-12-015.tex 222084 2013-12-30 14:48:41Z alverson $
\newlength\cmsFigWidth
\ifthenelse{\boolean{cms@external}}{\setlength\cmsFigWidth{0.85\columnwidth}}{\setlength\cmsFigWidth{0.49\textwidth}}
\ifthenelse{\boolean{cms@external}}{\providecommand{\cmsLeft}{top}}{\providecommand{\cmsLeft}{left}}
\ifthenelse{\boolean{cms@external}}{\providecommand{\cmsRight}{bottom}}{\providecommand{\cmsRight}{right}}

\providecommand{\cPQT}{\ensuremath{\mathrm{T}}\xspace}
\providecommand{\cPaQT}{\ensuremath{\overline{\mathrm{T}}}\xspace}
\providecommand{\TTbar}{\ensuremath{\cPQT\cPaQT}\xspace}
\providecommand{\bW}{\ensuremath{\cPqb\PW}\xspace}
\providecommand{\tZ}{\ensuremath{\cPqt\cPZ}\xspace}
\providecommand{\tH}{\ensuremath{\cPqt\PH}\xspace}
\newcommand{\hT}{\ensuremath{H_{\mathrm{T}}}\xspace}
\newcommand{\sT}{\ensuremath{S_{\mathrm{T}}}\xspace}
\providecommand{\FASTJET}{\textsc{fastjet}\xspace}
\cmsNoteHeader{B2G-12-015} 
\title{Inclusive search for a vector-like T quark with charge $\frac{2}{3}$ in pp collisions at $\sqrt{s}=8$\TeV}

\date{\today}

\abstract{A search is performed for a massive new vector-like quark T, with charge $\frac{2}{3}$, that is pair produced together with its antiparticle in proton-proton collisions.
The data were collected by the CMS experiment at the Large Hadron Collider in 2012 at $\sqrt{s} = 8$\TeV and correspond to an integrated luminosity of  19.5\fbinv.
The T quark is assumed  to decay into three different final states, \bW, \tZ, and \tH.
The search is carried out using events with at least one isolated lepton.
No deviations from standard model expectations are observed, and lower limits are set on the T quark mass
at 95\% confidence level. The lower limit lies between 687 and 782\GeV for
all possible values of the branching fractions into the three different
final states assuming strong production. These limits are the most stringent constraints to date on the existence of such a quark.}

\hypersetup{%
pdfauthor={CMS Collaboration},%
pdftitle={Inclusive search for a vector-like T quark with charge 2/3 in pp collisions at sqrt(s)=8 TeV},%
pdfsubject={CMS},%
pdfkeywords={CMS, physics}}

\maketitle 
\section{Introduction}

The discovery of a Higgs boson with a mass close to 125\GeV, with properties consistent with those of a standard model (SM) Higgs particle~\cite{Aad:2012tfa,Chatrchyan:2012ufa, Chatrchyan:2013lba}, suggests the need for a mechanism to stabilize the mass of this particle. Loop corrections to the mass of a scalar particle diverge quadratically with the cutoff scale of the calculation. The dominant contributions arise from loops that involve top quarks, \PW\ bosons, and Higgs bosons. If the SM applies to energies significantly above the electroweak scale, there must be other new particles that give rise to loop corrections that cancel these contributions. Little Higgs models~\cite{ArkaniHamed:2001nc, Schmaltz:2005ky}, for example, predict a quark ``T'', a partner to the top quark, which would cancel the contributions of the top-quark loops to the Higgs-boson mass. This T quark must have a mass at the TeV scale if it is to effectively fulfill this role.
Here we assume that the T quark is vector-like, i.e.\ that it has only vector couplings with the \PW\ and \Z bosons,  thereby evading the many constraints placed by precision electroweak measurements~\cite{LEP-2} on extensions to the SM that propose a fourth generation of quarks and leptons.

We assume that the T quark is  produced together with its antiquark in
proton-proton (pp) collisions through the strong interaction. Thus its production cross section can be calculated using perturbative quantum chromodynamics. We use the approximate next-to-next-to-leading order (NNLO) calculation implemented in \textsc{hathor}~\cite{HATHOR}, which gives results varying from 570\unit{fb} to 0.05\unit{fb} for T-quark masses between 500\GeV and 1500\GeV. A recent exact NNLO calculation~\cite{Czakon:2013goa} gives consistent results. The T quark can decay into three different final states: \bW, \tZ, or \tH.
At low T-quark masses, the \tZ and \tH modes are kinematically suppressed.
If the T quark is assumed to be an electroweak singlet, the branching fractions
should be approximately 50\% into \bW and 25\% each into \tZ and \tH when using the Goldstone Equivalence assumption~\cite{delAguila:1989rq}.
We will call these the nominal branching fractions.

We search for a T-quark signal without making any specific assumptions on the branching fractions. This is the first search that considers all three final states. Previous searches have considered a single final state or two final states.
The Compact Muon Solenoid (CMS) collaboration excluded T quarks that decay 100\% into \tZ for masses below 625\GeV~\cite{Chatrchyan:2011ay}.
T quarks that decay 100\% into \bW were excluded for masses below 570\GeV~\cite{Chatrchyan2012307,Chatrchyan2012103} and for masses below 656\GeV~\cite{Aad20131284} by the CMS and ATLAS collaborations, respectively.

All three decay channels produce final states with \cPqb\ quarks and \PW\ bosons. Here, we consider final states in which at least one \PW\ boson decays leptonically.

\section {The CMS detector}

The characteristic feature of the CMS detector is a superconducting
solenoid, 6\unit{m} in diameter and 13\unit{m} in length, which provides an axial
magnetic field of 3.8\unit{T}.
CMS  uses a right-handed cartesian coordinate system with its origin at the center of the detector. The $z$ axis coincides with the axis of symmetry of the detector, and is oriented in
the counterclockwise proton beam direction. The $x$ axis points towards the center of the Large Hadron Collider (LHC) ring.
The polar angle $\theta$ is defined with respect to the positive $z$
axis and $\phi$ is the corresponding azimuthal angle.
Pseudorapidity is defined as ${\eta = -\ln[\tan(\theta/2)]}$.

Several particle detection systems are located within the bore of the solenoid.
A multi-layered silicon pixel and strip tracker covering the
pseudorapidity region $\abs{\eta} < 2.5$  measures the trajectories of
charged particles. An electromagnetic calorimeter (ECAL) covering
$\abs{\eta}<3.0$ made of lead tungstate crystals, with a lead scintillator preshower detector covering $1.65 < \abs{\eta} < 2.6$, measures electrons and photons.
A hadron calorimeter made of brass and scintillators covering
$\abs{\eta} < 3.0$ measures jets. Muons are measured with gas-ionization detectors covering $\abs{\eta} < 2.4$
embedded in the steel flux return yoke of the solenoid,
and with the pixel and strip trackers.
The CMS detector is nearly hermetic, enabling momentum imbalance
measurements in the plane transverse to the beam directions.
A two-level trigger system selects the most interesting pp collision
events for use in physics analyses.
The Level-1 system uses custom hardware processors to select events in less than 4\mus, using information from the calorimeters and muon detectors. The high-level trigger processor farm further reduces the event rate to a few hundred Hz.
A detailed description of the CMS detector can be found in Ref.~\cite{cms}.

\section {Event samples}

The analysis is based on data recorded by the CMS experiment in pp collisions at $\sqrt{s} = 8$\TeV during the 2012 LHC run and corresponding to an integrated luminosity of 19.5\fbinv.
The inclusive muon sample is defined by the requirement to have an isolated muon candidate in the event with the transverse momentum $\pt >24$\GeV, as identified online by the trigger system. In the inclusive electron sample,  an isolated electron candidate in the event with $\pt >27$\GeV is required at the trigger level.
The multilepton sample consists of events with two or more isolated electron and/or muon candidates. At the trigger level, one lepton candidate must have $\pt>17$\GeV and the other $\pt>8$\GeV. The data are filtered to remove spurious events from noise or beam backgrounds by requiring a primary interaction vertex, and to remove data collected at times when the detector was not operating optimally.

The signal efficiencies and background contributions are estimated using simulated event samples.
The $\Pp\Pp\rightarrow\TTbar$ process is simulated using version 5.1.1 of the {\MADGRAPH}~\cite{madgraph5} event generator with up to two additional hard partons. For every T-quark mass between 500 and 1500\GeV, in 100\GeV increments, six different samples each with one of the possible final states ($\bW\bW$, $\bW\tH$, $\bW\tZ$, $\tH\tH$, $\tH\tZ$, and $\tZ\tZ$) are generated. All possible combinations of branching fractions can be simulated by combining these samples with the appropriate weights.
The Higgs boson decays are simulated assuming SM branching fractions for a mass of 125\GeV.

Events from SM processes that give rise to backgrounds are generated using {\MADGRAPH} ({\PW}+jets, {\Z}+jets, $\ttbar \PW$, and $\ttbar \Z$ production), \POWHEG version 1~\cite{powheg1, powheg2, powheg3} ($\ttbar$ and t production), and \PYTHIA version 6.424~\cite{pythia} (\PW\PW, \PW\Z, \cPZ\cPZ, and $\ttbar \PH$ production).
For W+jets and Z+jets production, {\MADGRAPH} generates samples with up to four partons. These samples are merged using the MLM scheme with $k_T$ jets~\cite{Alwall:2007fs,Alwall:2008qv}. For \POWHEG the CTEQ6M parton distribution functions (PDFs) are used and for all other generators the CTEQ6L1~\cite{1126-6708-2002-07-012} PDFs are used.  Hadronization and parton showering are simulated using \PYTHIA for all samples, and the CMS detector response is simulated using \GEANTfour~\cite{geant4}.
Minimum bias interactions, generated using \PYTHIA, are superimposed on the simulated events to model the effect of additional pp collisions within a single bunch crossing (pileup).
The simulated interaction multiplicities are made to match the data, given the observed luminosity profile. The average number of simultaneous collisions per bunch crossing in the data sample is 21. The normalization of the {\PW}+jets sample is determined directly from the data, and all other samples are normalized to the next-to-leading-order prediction of their cross sections as computed with \textsc{mcfm}~\cite{MCFM}.

\section{Event reconstruction}

The event vertex of the hard scatter, ``primary vertex'', is identified as the reconstructed vertex with the largest $\sum\pt^2$ of its associated tracks.
Data and simulated samples are reconstructed by a particle-flow algorithm~\cite{CMS-PAS-PFT-10-002}, which reconstructs all visible particles in the event originating from the primary interaction.
Charged particles identified as coming from pileup interactions are not considered.

Muon candidates~\cite{Chatrchyan:2012xi} are reconstructed from track segments detected in the muon chambers combined with matching hits in the silicon tracker. Electron candidates~\cite{CMS-PAS-EGM-10-004, Chatrchyan:2013dga} are reconstructed as clusters of energy deposits in the ECAL that are consistent with a track in the silicon tracker. Electron candidates consistent with arising from a photon conversion are rejected.
An isolation variable is defined as the ratio of the sum of \pt of all additional  particles reconstructed in an isolation cone to the  \pt of the lepton candidate. The cone radius is $\Delta R = \sqrt{\smash[b]{(\Delta\phi)^2+(\Delta\eta)^2}}=0.4$ around muon candidates and  $\Delta R =  0.3$ around electron candidates.
The sum of \pt in the isolation cone is corrected,  on an event-by-event basis, for the remaining contributions from other interactions in the same beam crossing. A muon is considered isolated if the isolation variable is below {0.12}.
For electrons the corresponding requirement is {0.10}.

All reconstructed particles except isolated leptons are clustered into jets using the anti-\kt jet clustering algorithm~\cite{antikt} with a distance parameter of 0.5, as implemented in {\FASTJET} {3.0}~\cite{Cacciari:2011ma}.  Energy response, trigger and reconstruction efficiencies for simulated event samples are corrected using scale factors determined from data to reproduce the performance of the CMS detector~\cite{1748-0221-6-11-P11002}. Efficiency corrections are of order a few percent.  Jet energy corrections vary between 1\% and 10\%, depending on $\eta$ and \pt.

The missing transverse energy, \MET, is defined as the magnitude of the  vector sum of the transverse momenta of all reconstructed particles.
We define \hT as the scalar sum of the transverse momentum of all jets, and \sT as the sum of \hT, \MET, and the  transverse momenta of all leptons.

Jets originating from the
hadronization of a b quark are identified by the combined secondary vertex
algorithm~\cite{Chatrchyan:2012jua}, which combines information
about impact parameter significance, secondary-vertex reconstruction, and jet kinematic properties. Jets identified by the algorithm are said to be \cPqb-tagged. For jet kinematics typical of top-quark decays, the algorithm has a $66.1\pm0.3$\% probability of tagging jets from b quarks and a $1.3\pm0.2$\% probability of tagging jets from light quarks and gluons~\cite{CMS-PAS-BTV-13-001}.

For large values of the T-quark mass, its decay products have large
\pt values and their secondary decay products may get merged into a
single jet.
In order to identify highly boosted \PW-boson and top-quark jets from the decay of massive particles, we perform an additional jet reconstruction using the Cambridge--Aachen algorithm~\cite{1126-6708-1997-08-001} with a distance parameter of {0.8}.
Jets with $\pt>200$\GeV and a mass between 60 and 130\GeV are classified as \PW\ jets~\cite{boostedhiggs,jetpruning1,jetpruning2}.
This signature is most important for T decays to \bW because in this decay
the \PW\ boson tends to have the largest \pt. It can also occur in T decays to \tZ or \tH but here the decay products of the bosons merge less often
because in these decays the boson is accompanied by the massive top quark
and therefore has smaller \pt. The decay products of a hadronic
top decay may merge into a single jet. To identify top-quark jets, we
follow the method of Ref.~\cite{Chatrchyan:2012ku}.
Jets are classified as top jets if they have $\pt>200$\GeV, a mass between 140 and 250\GeV, at least three subjets, and the minimum pairwise mass larger than 50\GeV.
The efficiency for identifying \PW\ and top jets is adjusted
for differences in the range of 5-7\% between data and simulation.
The \PW\ and top jet reconstruction is independent of the standard jet reconstruction, and the collection of jets reconstructed by the latter is not modified by the \PW\ and top jet identified.

\section {Single-lepton channel}

Single-muon events are selected in the inclusive muon sample requiring an isolated muon candidate with $\pt>32$\GeV and $\abs{\eta}<2.1$.
Single-electron events are selected in the inclusive electron sample requiring an isolated electron candidate with $\pt>32$\GeV and $\abs{\eta}<1.44$ or $1.57<\abs{\eta}<2.5$.
In each case, the candidate lepton must be consistent with originating from the primary vertex. Events that have a second muon or electron candidate are removed from the sample. All events must have at least three jets with $\pt > 120$, 90, and~50\GeV respectively. In addition, at least one \PW\ jet has to be identified or there has to be a fourth jet with $\pt > 35$\GeV. Each of the jets must have $\abs{\eta} < 2.4$ and be separated by $\Delta R > 0.4$ from the isolated muon and by $\Delta R > 0.3$ from the isolated electron. Requiring several high-\pt jets greatly reduces the contributions from SM background processes, which are all dominantly produced with fewer and softer jets. All events must also have $\MET > 20$\GeV. Combined with the requirements above, this last requirement effectively suppresses contributions from background multijet events.

To avoid large uncertainties from modeling \PW-boson production in association with multiple energetic jets, the \PW-boson background is normalized directly to a control data sample consisting of events selected in exactly the same way as in the signal selection but with the requirement that the events have at most three jets with $\pt > 35$\GeV and no \PW\ jet. This sample is dominated by \PW-boson and top-quark production, and would have a negligible signal contribution. We determine two scale factors such that the total number of simulated events and the number of simulated events with at least one \cPqb-tagged jet agree with the corresponding counts observed in the control data sample. One scale factor is used to multiply the number of events with a \PW\ boson and heavy-flavor (\cPqb- or \cPqc-quark) jets; the other scale factor is used to multiply the number of events with a \PW\ boson without heavy-flavor jets.  In addition, we scale events containing \cPqb- and \cPqc-quark jets with two different scale factors.
The ratio of these two scale factors is set to the value determined
in the semileptonic $\ttbar$ sample at $\sqrt{s}=7$\TeV from~\cite{Chatrchyan201383}. The scale factors are 0.8 for events that have at least one \cPqb\ quark, 1.1 for events without \cPqb\ quarks but at least one \cPqc\ quark, and 1.0 for events with only light quarks and gluons. These factors are applied after the samples are normalized to the inclusive \PW-boson production cross section predicted at NNLO~\cite{Melnikov:2006kv}.
The same scale factors are applied to events with electrons and to events with muons. 

\begin{figure*}[hbtp]
\begin{center}
\includegraphics[width=0.48\textwidth]{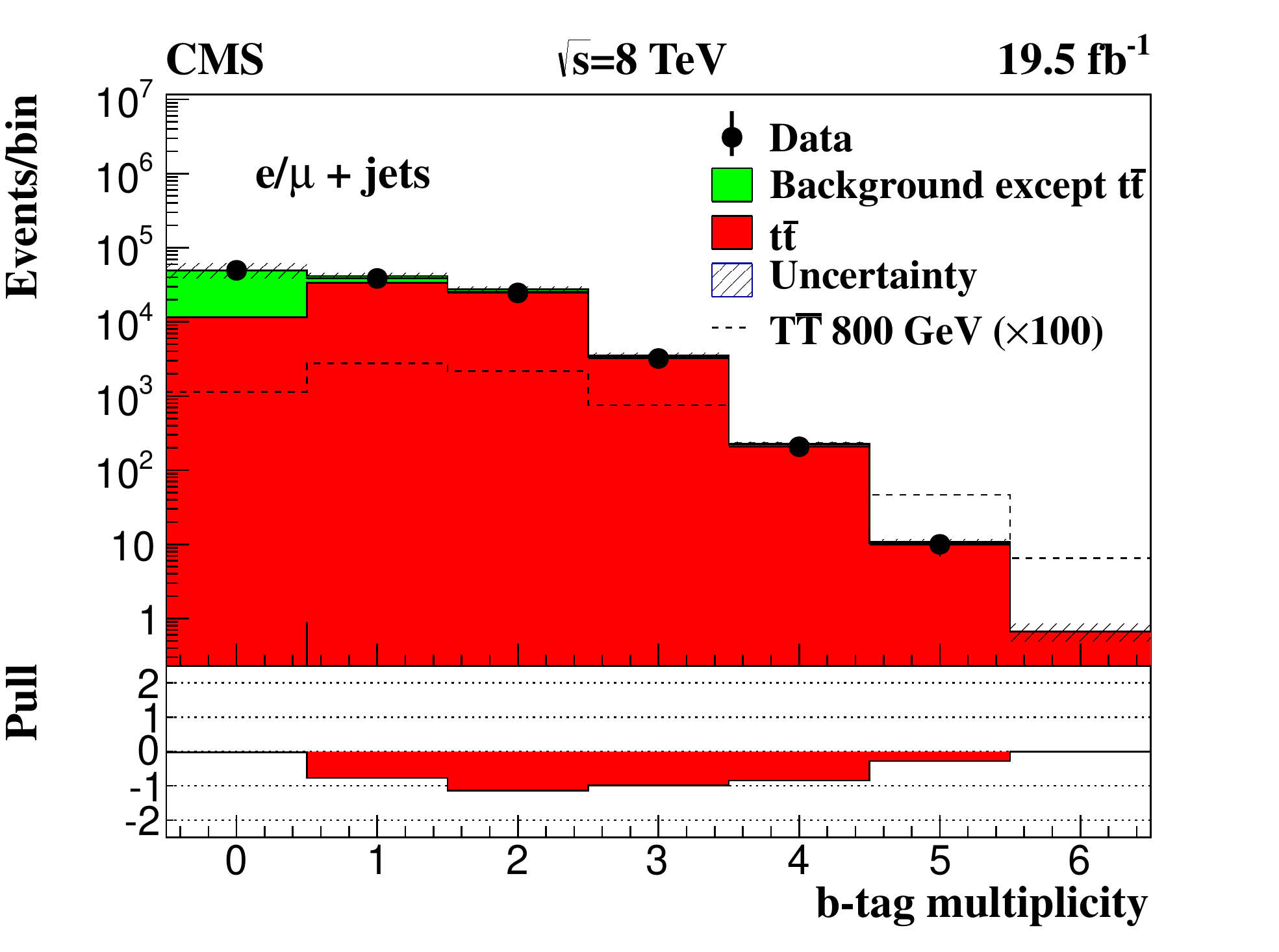}
\includegraphics[width=0.48\textwidth]{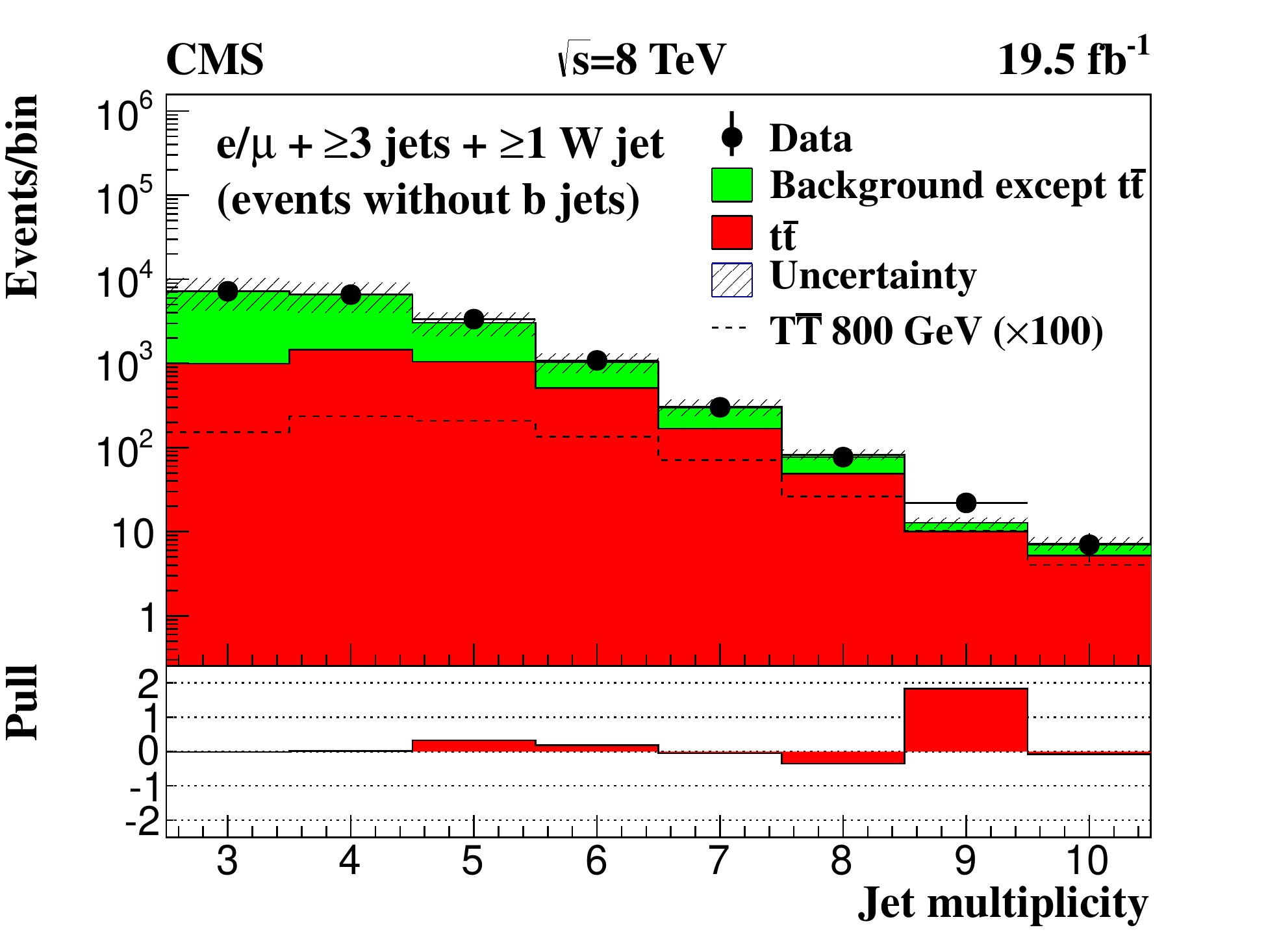}
\caption{Observed multiplicity of \cPqb-tagged jets in the single-lepton sample compared with a simulation using the \PW-boson background normalization determined from the data (left) and observed multiplicity of jets with $\pt>30$\GeV for events with one isolated lepton, at least three jets, at least one W jet and no \cPqb-tagged jets (right).
The bin-by-bin pulls shown in this and other figures are the values of the difference
between observed number and expected number of events divided by the sum in quadrature of the systematic and statistical uncertainties.
The uncertainties are correlated bin-to-bin, and include those in the luminosity,  the cross sections and the correction factors, as described in Section~\ref{sec:limit}.
}
\label{fig:nbtags}
\end{center}
\end{figure*}

Figure~\ref{fig:nbtags} shows that the overall jet multiplicity distribution and the multiplicity of b-tagged jets are both well modeled by the simulation following the scaling procedure. The left plot in Fig.~\ref{fig:nbtags} shows the agreement between data and simulation for the multiplicity of b-tagged jets. As an additional cross-check of the simulation of the background we have looked at the overall jet multiplicity, in a subset of the event sample without b-tagged jets. This distribution is shown as the right plot in Fig.~\ref{fig:nbtags}.

The numbers of events expected and observed are given in Table~\ref{tab:yields}. The selection efficiencies and expected numbers of events for the T-quark signal, assuming nominal branching fractions, are summarized in Table~\ref{tab:efficiency}.

\begin{table}[htb]
\begin{center}
\topcaption{Number of events predicted for background processes and observed in the
single-lepton sample. The uncertainty in the total background expectation is computed including the correlations between the systematic uncertainties of the individual contributions. The uncertainties include those in the luminosity,  the cross sections and the correction factors, as detailed in Section~\ref{sec:limit}.
}\label{tab:yields}

\begin{tabular}{lr@{$\pm$}lr@{$\pm$}l}
\hline
Lepton flavor   	&\multicolumn{2}{c}{Muon} & \multicolumn{2}{c}{Electron} \\
\hline
$\ttbar$            	& 36700 	        & 5500   	    & 35900 	& 5400 \\
Single top quark       	& 2200          	& 1100           & 2100 	& 1000 \\
\PW                  	& 19700 	        & 9900         & 18600 	& 9400 \\
\Z                  	& 2200 	        & 1100     	    & 2000  	& 1000 \\
Multijets        	& \multicolumn{1}{r}{${<}60$}& & 1680	        & 620\\
$\ttbar \PW$        	& 144 	        & 72		    & 137 	        & 68 \\
$\ttbar \Z$        	& 109 	        & 54       	    & 108 	        & 54 \\
$\ttbar \PH$        	& 570               & 290          & 570 	        & 290 \\
\PW\PW/\PW\cPZ/\cPZ\cPZ  	& 410 	        & 200          & 400 	        & 200 \\
\hline
Total background & 61900 	        & 13900        & 61500 	& 13700 \\
Data           	& \multicolumn{1}{r}{58478}  &	 	& \multicolumn{1}{r}{57743}  \\
\hline
\end{tabular}
\end{center}
\end{table}

\begin{table*}[htb]
\begin{center}
\topcaption{Production cross section, efficiency, and number of events predicted by the single-lepton analysis, for the T-quark signal processes, assuming the nominal branching fractions into \bW, \tH, \tZ of 50\%, 25\%, 25\%, respectively.}\label{tab:efficiency}
\begin{tabular}{cccccc}
\hline
Lepton flavor    & Cross section&\multicolumn{2}{c}{Muon} &\multicolumn{2}{c}{Electron} \\
T mass (\GeVns{}) &  (fb) & Efficiency & Events & Efficiency & Events \\
\hline
~500 &	571 	&	7.6\% &	850	&	7.5\% &	840\\
~600 &	170	&	8.3\% &	280	&	8.4\% &	280\\
~700 &	56.9 	&	8.7\% &	97	&	8.8\% &	98\\
~800 &	20.8 	&	8.9\% &	36	&	9.1\% &	37\\
~900 &	8.09 	&	9.0\% &	14.3	&	9.3\% &	14.8\\
1000 &	3.27 	&	9.0\% &	5.8	&	9.4\% &	6.0	\\
1100 &	1.37 	&	9.0\% &	2.4	&	9.4\% &	2.5	\\
1200 &	0.58 	&	9.0\% &	1.0	&	9.4\% &	1.1	\\
\hline
\end{tabular}
\end{center}
\end{table*}

\begin{figure*}[hbt]
\begin{center}
\includegraphics[width=0.49\textwidth]{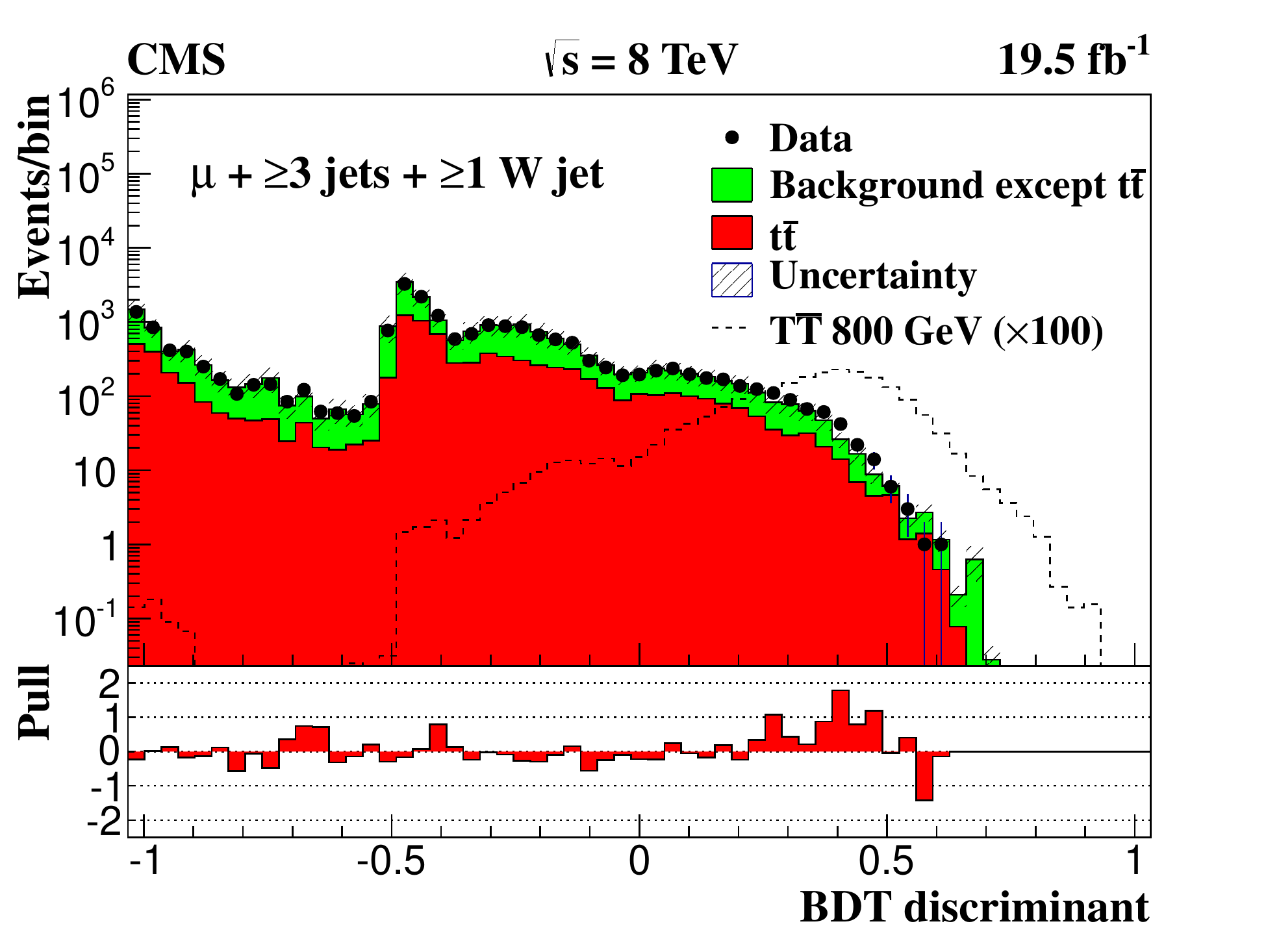}
\includegraphics[width=0.49\textwidth]{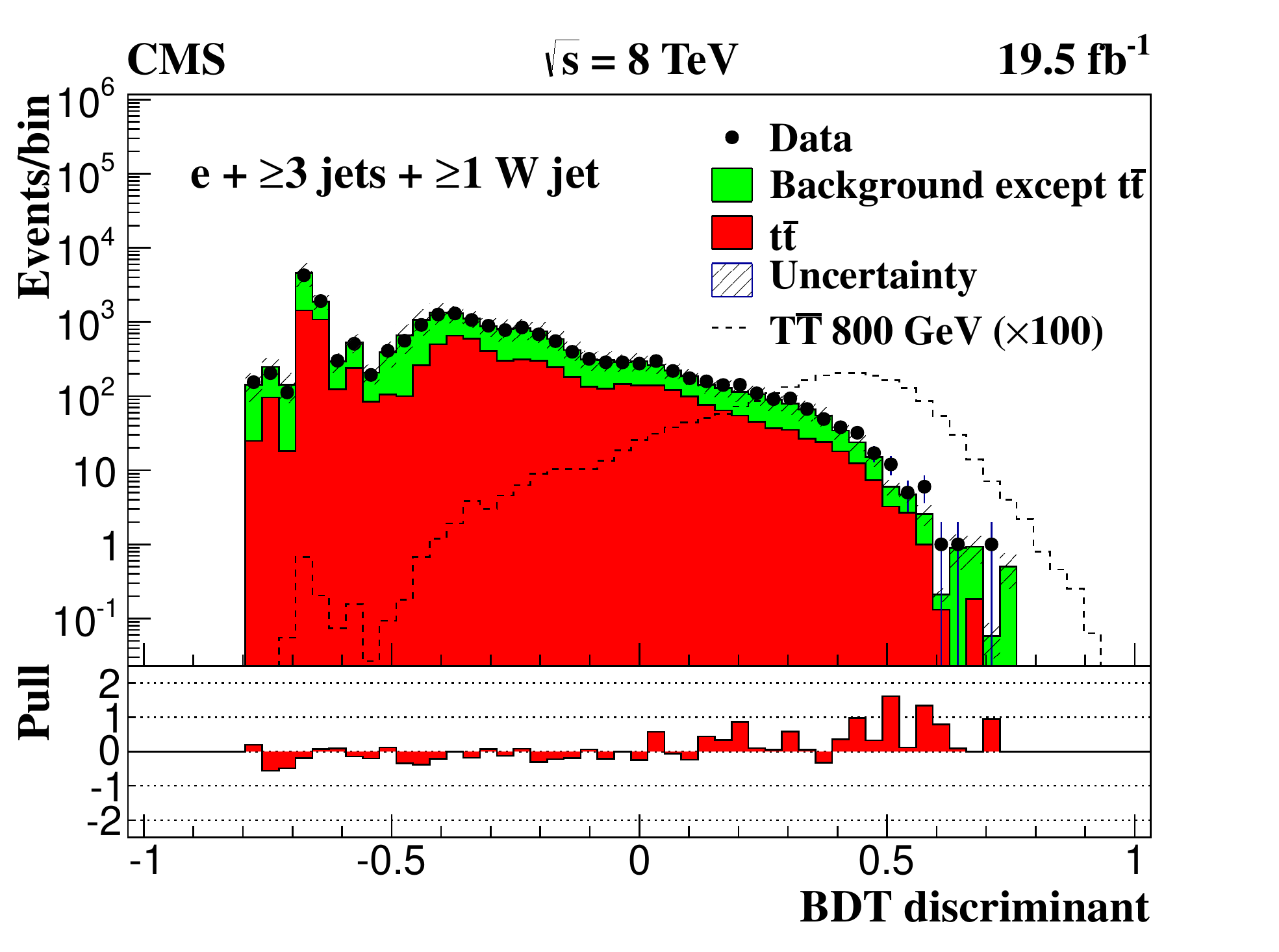}
\includegraphics[width=0.49\textwidth]{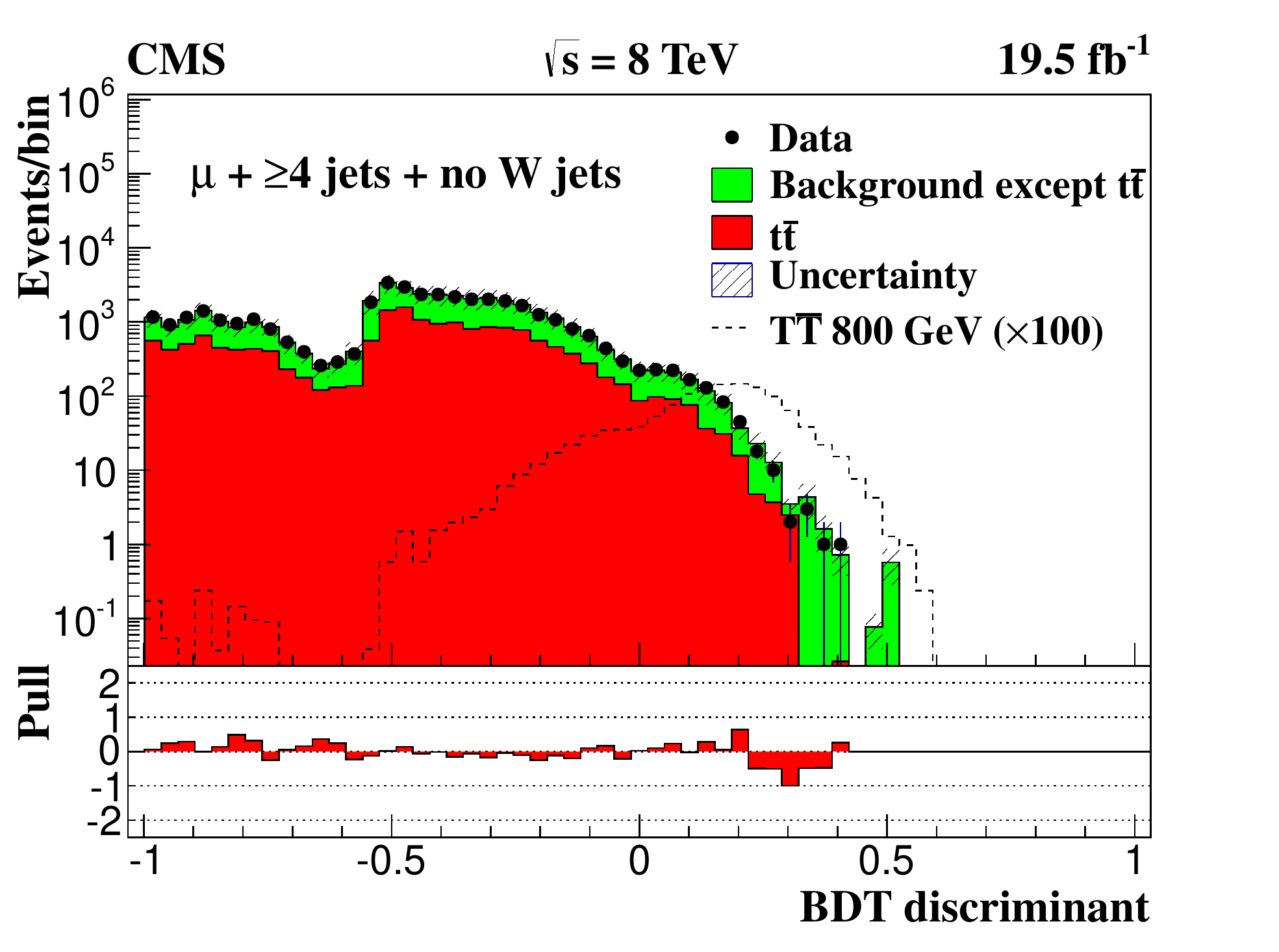}
\includegraphics[width=0.49\textwidth]{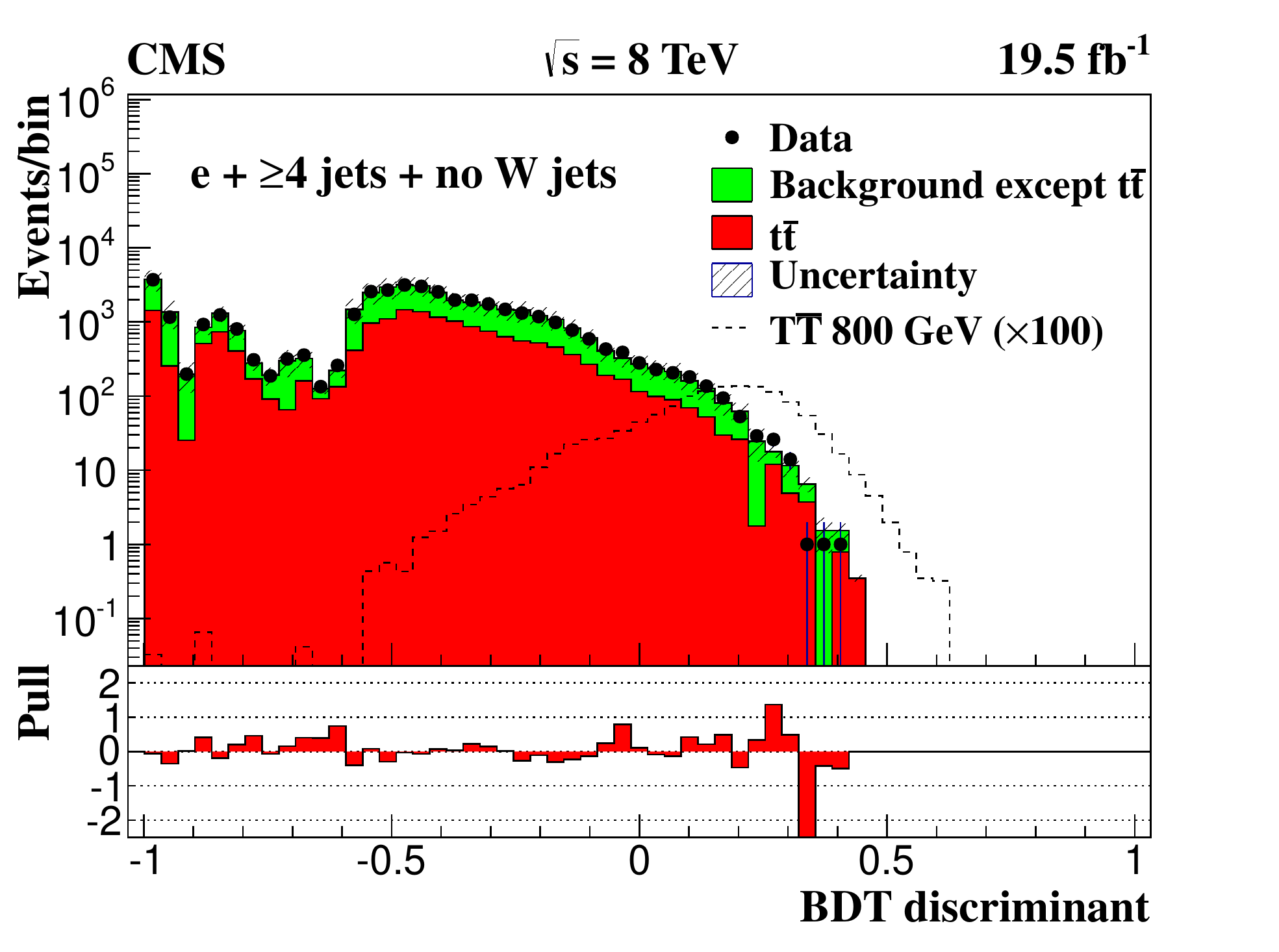}
\caption{
Observed and expected distributions of the BDT discriminant.
The distribution for a T quark with a mass of 800\GeV is also shown. The top panel is for events with at least one \PW\ jet, the bottom panel for events without \PW\ jets.
The left column is for events with a muon and the right column for events with an electron.}
\label{fig:BDT}
\end{center}
\end{figure*}

We use boosted decision trees (BDT)~\cite{Hocker:2007ht} to further separate the T-quark signal from the SM background, more than  96\% of which arises from $\ttbar$, \PW- and \Z-boson production. In the training of the BDT, we include the signal sample with the composition defined by the nominal branching fractions.
We have tried training separate BDTs using T quark samples decaying 100\% to one of the three final states \bW, \tZ, or \tH.
This procedure did not lead to a significant improvement in sensitivity and therefore we use the same BDT for all combinations of branching fractions. Only  $\ttbar$, \PW- and \Z-boson production contributions enter the BDT training. We train separate BDTs for events with at least one \PW\ jet and for events without any \PW\ jet, at every value of the T-quark mass.
The BDT distributions for the T-quark signal move towards slightly higher values and get a little wider with increasing mass.
Although our sample includes all SM decays of the Higgs boson, we are mostly sensitive to decays to \cPqb-quark pairs and vector bosons with hadronic decays.
We split the signal and background samples into two subsamples and use one of the subsamples to train the BDT and the other to model the BDT discriminant distribution to be compared with the data.
The input variables for the BDT are jet multiplicity, \cPqb-tagged jet multiplicity, \hT, \MET, lepton \pt, \pt of the third jet, and \pt of the fourth jet. For events with a \PW\ jet, the number and \pt of \PW\ jets and the number of top jets are included as additional parameters. These variables are chosen based on their importance calculated by the BDT algorithm and the desire to avoid strong correlations between the input variables. We have verified that the distributions of these variables agree well with expectations. The distributions of the BDT discriminant are shown in Fig.~\ref{fig:BDT}.
These demonstrate the discrimination between the T-quark signal and the SM background.

\begin{figure*}[hbt]
\begin{center}
\includegraphics[width=0.49\textwidth]{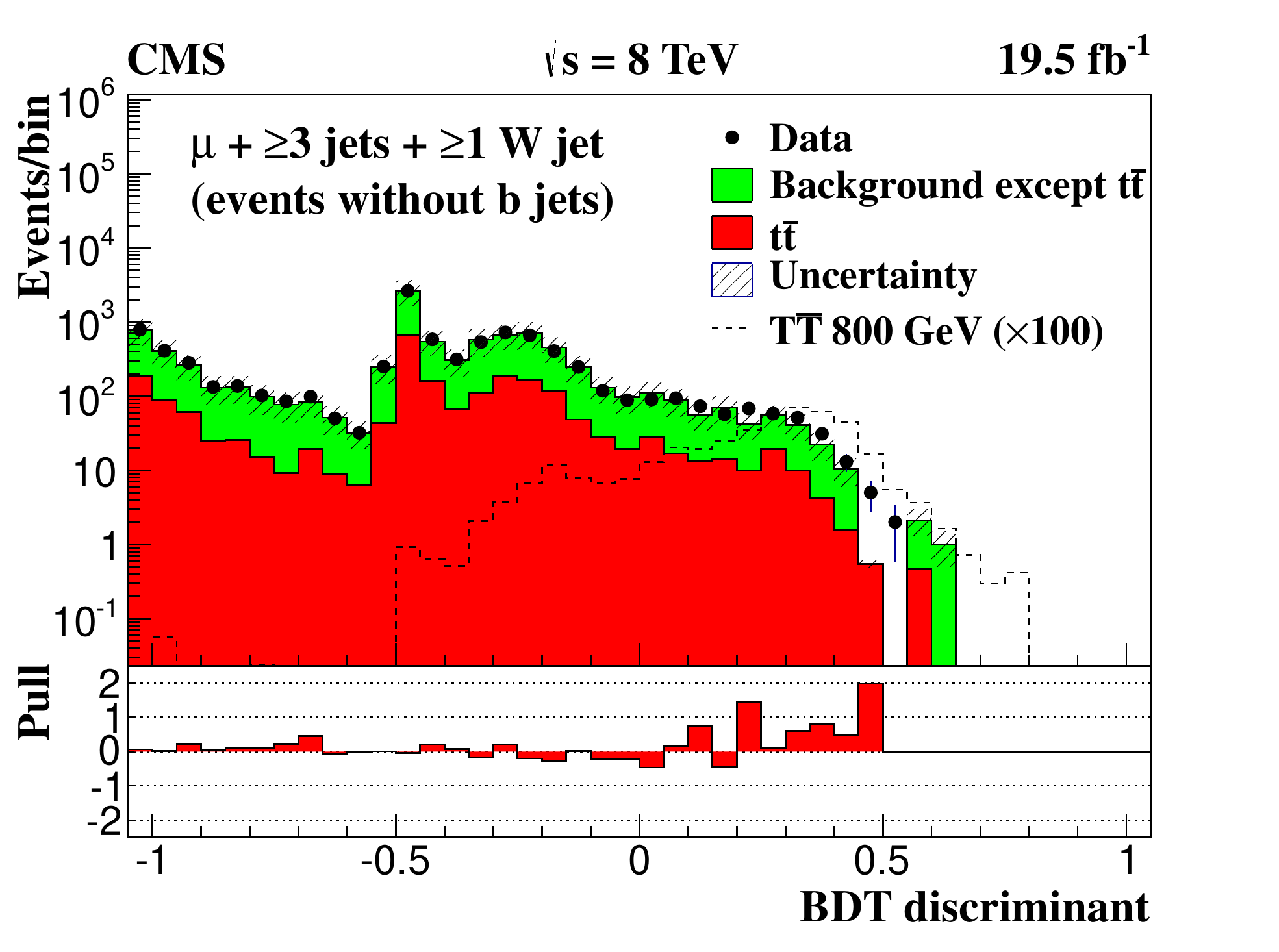}
\includegraphics[width=0.49\textwidth]{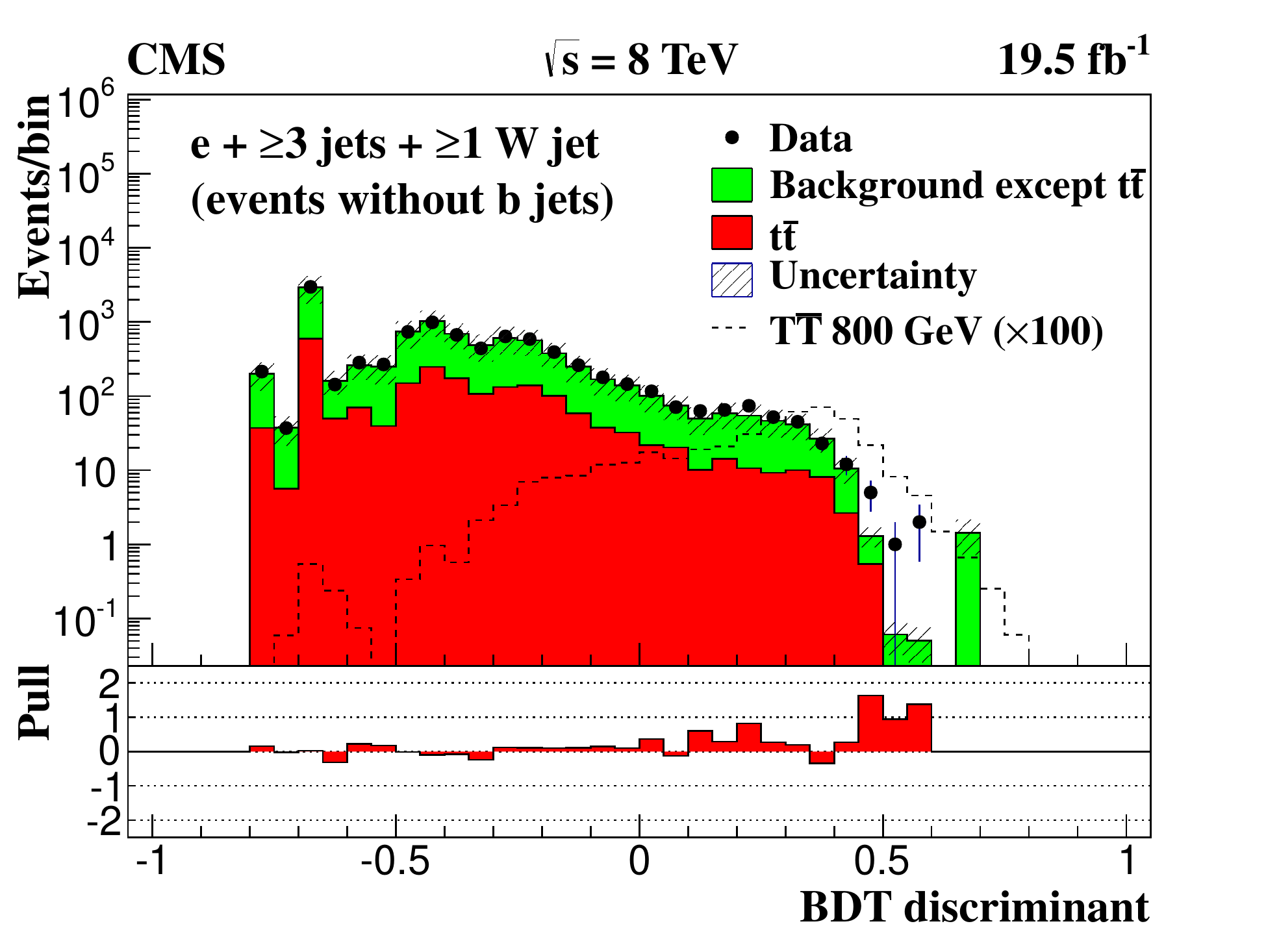}
\includegraphics[width=0.49\textwidth]{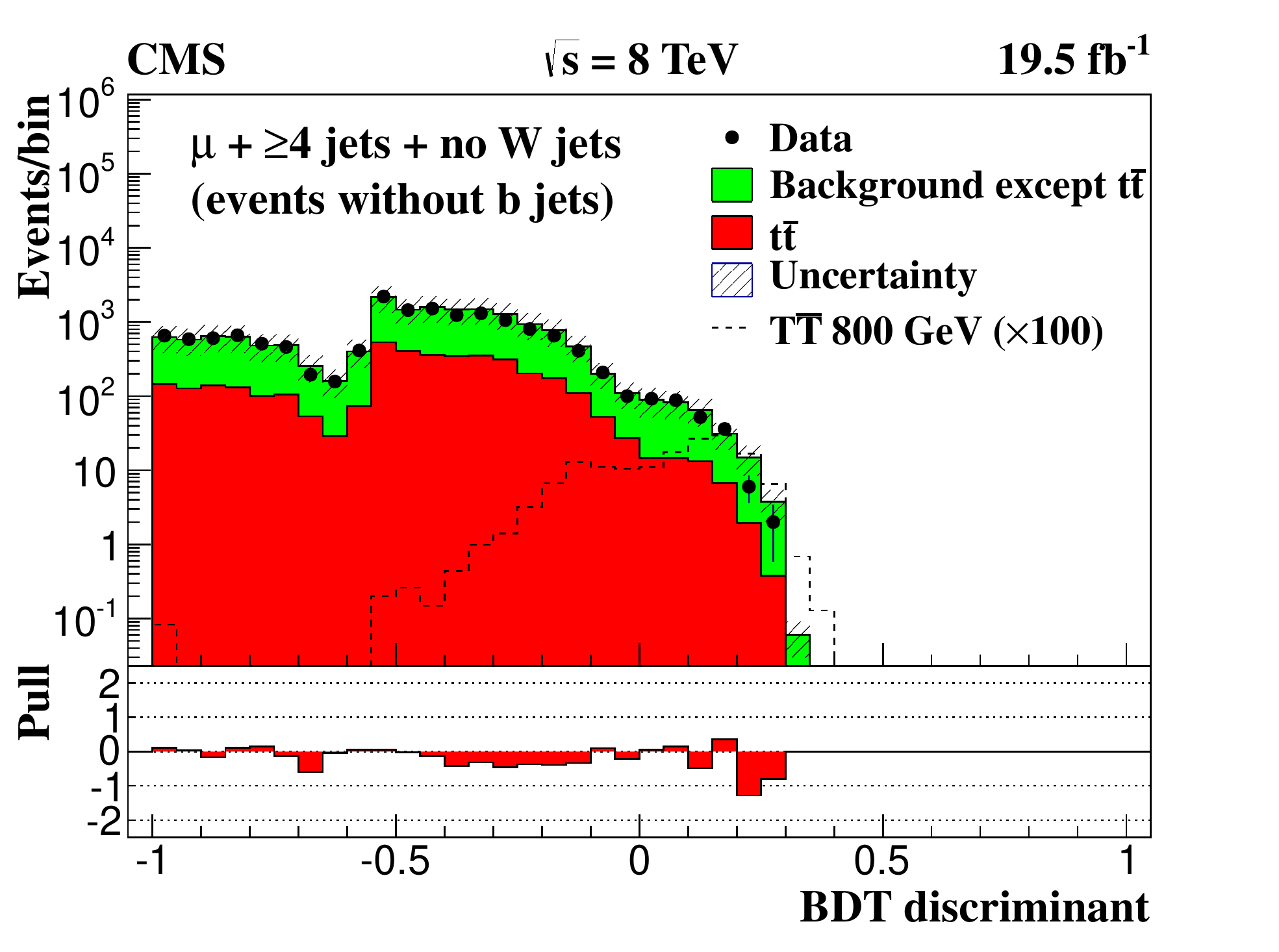}
\includegraphics[width=0.49\textwidth]{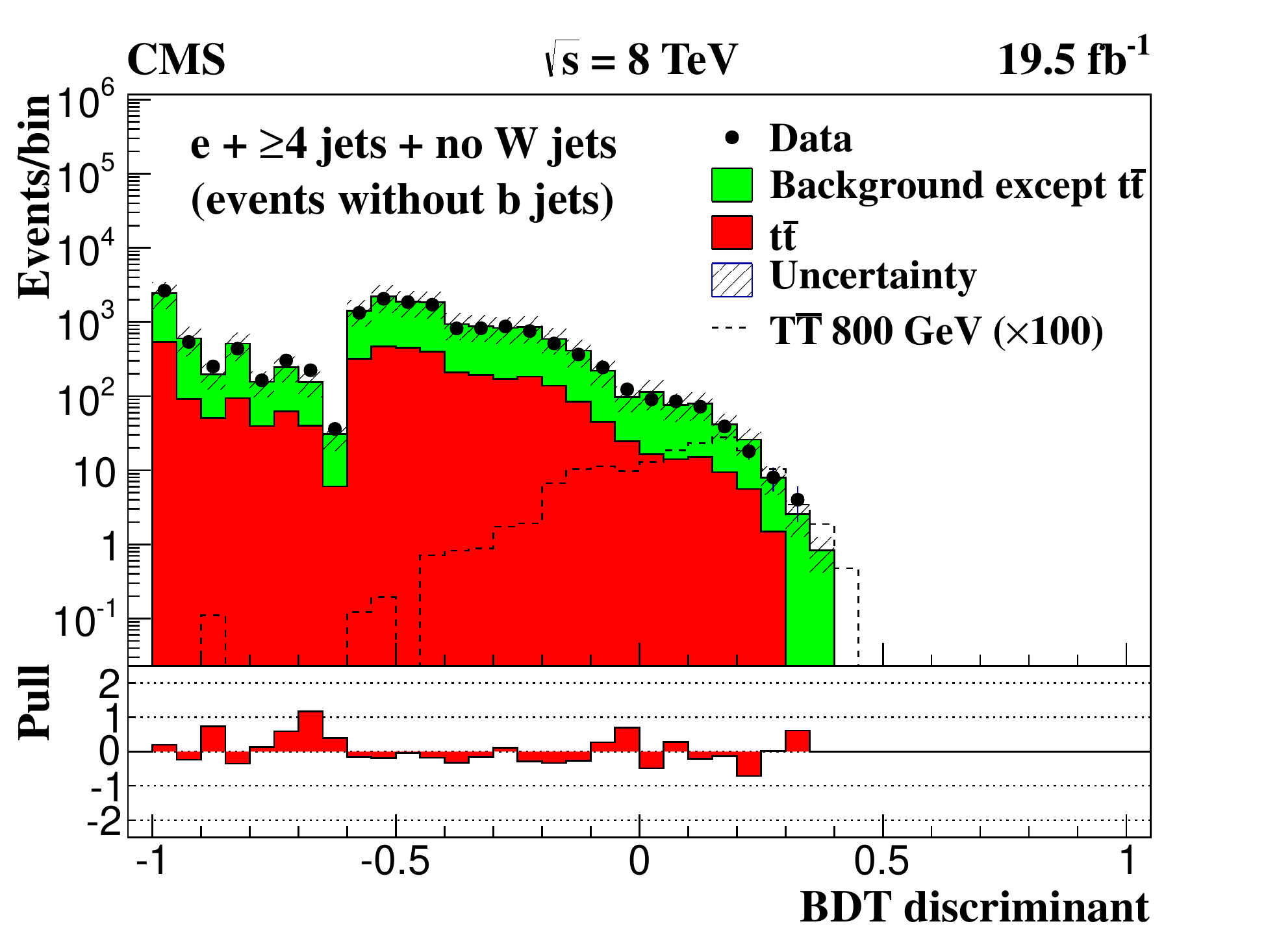}
\caption{
Observed and expected distributions of the BDT discriminant for the subset of events in the subsample without \cPqb-tagged jets.
The distribution for a T quark with a mass of 800\GeV is also shown. The top panel is for events with at least one \PW\ jet, the bottom panel for events without \PW\ jets.
The left column is for events with a muon and the right column for events with an electron.}
\label{fig:BDTcheck}
\end{center}
\end{figure*}

As an auxiliary check, we show that the simulation models the data well by comparing the distributions of the BDT discriminant in the subset of the sample without \cPqb-tagged jets as shown in Fig.~\ref{fig:BDTcheck}.
In this sample the signal is suppressed by a factor 5 relative to the default selection with W jets and by a factor 8 for the sample without \PW\ jets.

\section {Multilepton channel}

The multilepton sample is divided into the four mutually exclusive subsamples described below. Dilepton events are required to have exactly two leptons with $\pt>20$\GeV. These are divided into opposite- and same-sign dilepton events according to their charges, and the opposite-sign  sample is further divided in two samples according to the number of jets in the event. Trilepton events must have at least three leptons with $\pt>20$\GeV.
To reject heavy-flavor resonances and low-mass Drell--Yan (DY) production, we require at least one  dilepton pair with a mass above 20\GeV and $\MET>30\GeV$ in these samples.
Jets must have $\pt > 30\GeV$ and $\abs{\eta} < 2.4$ and be separated by $\Delta R > 0.3$ from the selected leptons.
We also require that at least one jet must be identified as a b jet.

\begin{figure}[hbt]
\begin{center}
\includegraphics[width=0.48\textwidth]{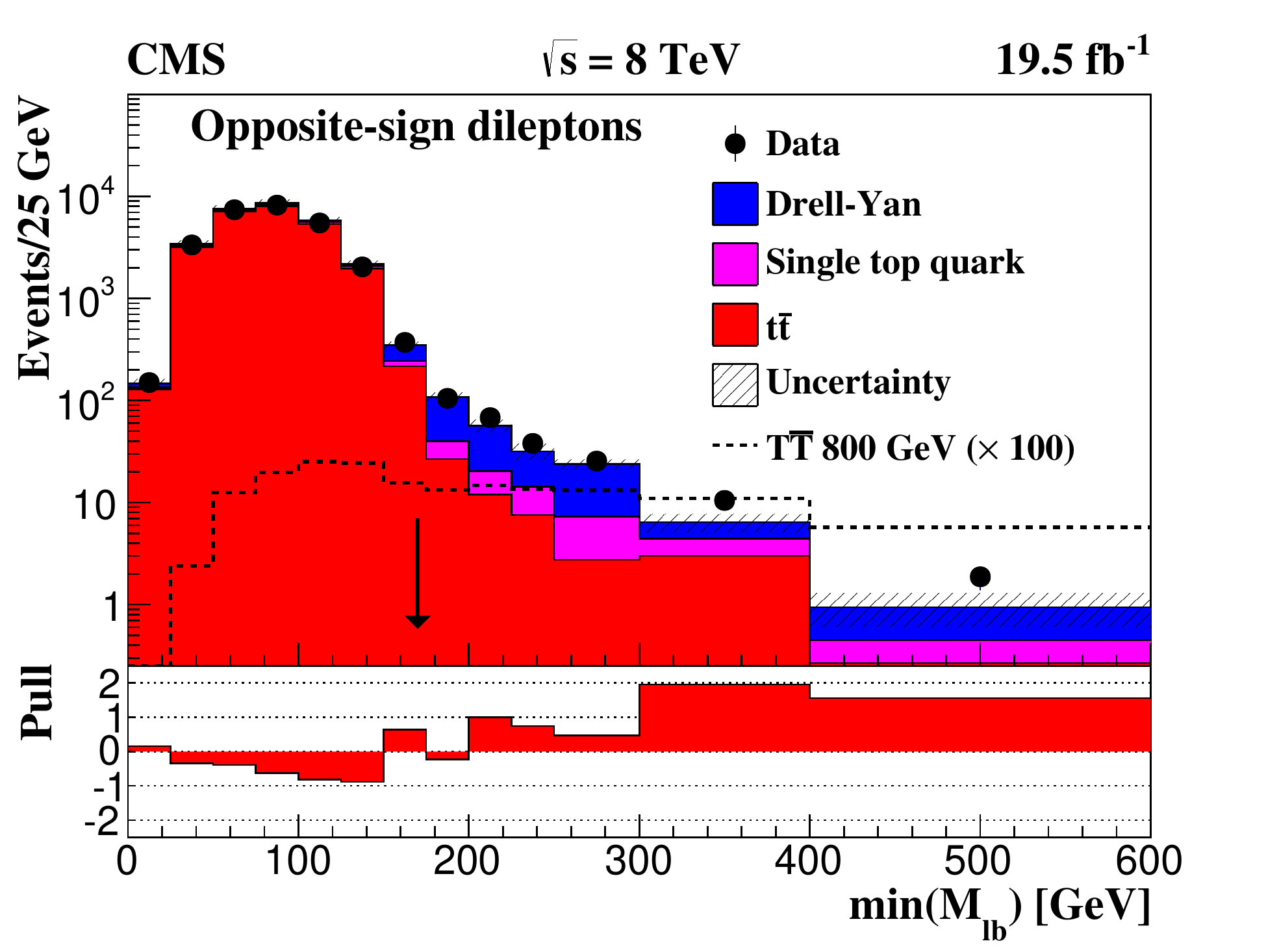}
\caption{Observed and expected distributions of the smallest $M_{\ell\cPqb}$ for the opposite-sign dilepton sample.
The distribution for a T quark with a mass of 800\GeV is also shown. It is dominated by the $\bW\bW$ final state.
The arrow indicates the chosen requirement.}
\label{fig:mlb}
\end{center}
\end{figure}

The first opposite-sign dilepton sample (referred to as the  \textit{OS1} sample) mostly accepts events in which both the T and the $\cPaQT$ quarks decay to \bW, resulting in a $\bW\bW$ final state~\cite{Chatrchyan2012103}. The main irreducible backgrounds in this sample are \ttbar and DY production. To minimize these backgrounds, we impose the following requirements. The mass of the dilepton pair, $M_{\ell\ell}$, must not be consistent with the \Z-boson mass, i.e.\ we eliminate events in which $76<M_{\ell\ell}<106$\GeV.
We require that the smallest invariant mass of lepton and \cPqb-jet
combinations, $M_{\ell\cPqb}$, is larger than 170\GeV. Since, in a top-quark decay,
$M_{\ell\cPqb}$ must be smaller than the top-quark mass, this drastically reduces the \ttbar\ background as can be seen in Fig.~\ref{fig:mlb}.
Finally, the events must have either two or three jets, $\hT>300$\GeV, and $\sT>900$\GeV. The final selection requirements are optimized by computing expected limits on the T-quark mass.

The DY background is not modeled adequately at low invariant mass and in the presence of missing transverse energy. We therefore use data to measure the residual background in events with two muons or two electrons.
The observed event count in the \Z-boson mass peak is rescaled by the ratio of DY events outside and inside the mass window as measured in a control data sample consisting of events with no \cPqb-tagged jets, $\MET<10$\GeV, $\sT < 700\GeV$, and $\hT > 300\GeV$. Since contamination from non-DY backgrounds can still be present in the \Z-boson mass window, this contribution is subtracted using the $\Pe\mu$ channel scaled according to the event yields in the $\mu\mu$ and $\Pe\Pe$ channels.

Events in the second opposite-sign dilepton sample (referred to as the  \textit{OS2} sample) must have at least five jets, of which two must be \cPqb-tagged, $\hT>500$\GeV, and $\sT>1000$\GeV. This sample accepts final states in which both leptons come from the decay of a \Z boson but is not sensitive to the $\bW\bW$ final state. The dominant background in this channel is $\ttbar$ production.

\begin{figure}[hbt]
\begin{center}
\includegraphics[width=0.48\textwidth]{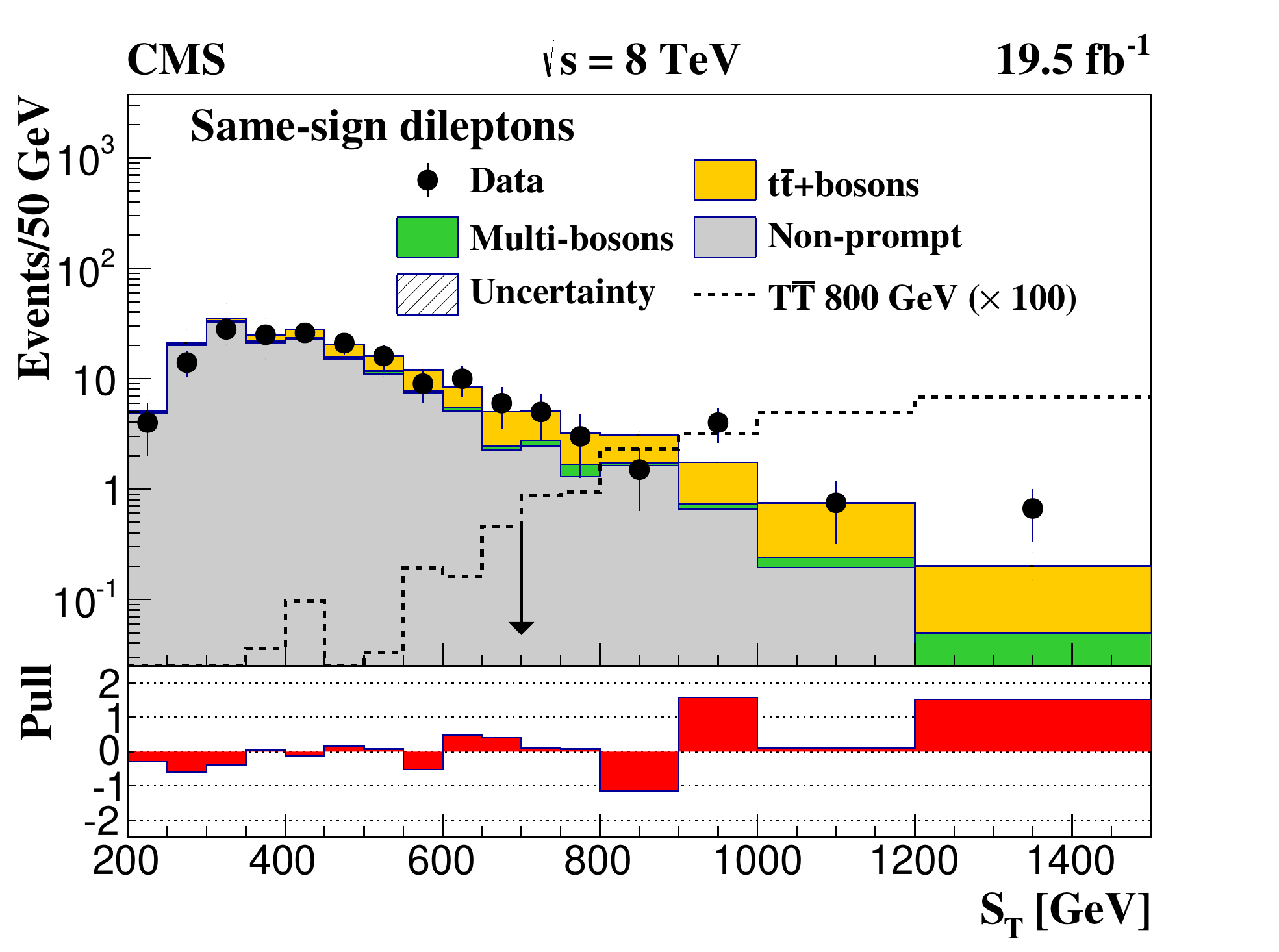}
\caption{Observed and expected distributions of \sT for the same-sign dilepton sample. The arrow indicates the chosen requirement.}
\label{fig:ST}
\end{center}
\end{figure}

The same-sign dilepton sample (the {\em SS} sample) accepts events in which at least one T quark decays to \tZ or \tH. The $\bW\bW$ final state does not contribute to this channel. We further filter these events by requiring at least three jets, $\hT>500$\GeV, and $\sT>700$\GeV. The distribution of \sT is shown in Fig.~\ref{fig:ST}.
The backgrounds associated with this channel fall into three main categories. Standard model processes leading to prompt, same-sign dilepton signatures have very small cross sections and are determined from simulation. Events with two prompt leptons of opposite charge can be selected if one lepton is misreconstructed with the wrong charge sign. The probability to misreconstruct the charge sign of a muon in the \pt range considered here is negligible. We determine the probability to misreconstruct the charge sign of an electron from a sample of \Z decays where events with oppositely charged leptons are selected with the same criteria as in the signal selection except for the charge requirement.
We then weight the events by the charge misreconstruction probability to determine the number of expected background events. The charge misidentification contribution to the background is dominated by events from $\ttbar$ production. We also determine instrumental backgrounds, where jet misidentification is the source of one or both lepton candidates, using control data samples.

The trilepton sample also accepts events in which at least one T quark decays to \tZ or \tH. The $\bW\bW$ final state does not contribute to this channel. We further filter trilepton events requiring at least three jets,  $\hT>500$\GeV, and $\sT>700$\GeV. The backgrounds in this channel originate from SM processes with three or more leptons in the final state, such as diboson and triboson production, which are modeled by simulation. There are also non-prompt backgrounds from $\ttbar$ production and other processes, characterized by one or more misidentified leptons. These are determined from data as for the dilepton samples.

The numbers of events expected and observed in the multilepton samples are given in Table~\ref{tab:yields_ll}. The selection efficiencies and expected numbers of events for the T-quark signal, assuming nominal branching fractions, are summarized in Table~\ref{tab:efficiency_ll}.
The selection efficiencies decrease for large values of the T-quark mass, above 1100\GeV,  because an increasing fraction of the decay products of \PW\ and \Z bosons are reconstructed as single jets.
For the multilepton samples, the numbers of events expected from background and the T-quark signal are of similar order of magnitude and therefore we use the event count in the different multilepton samples, distinguished by lepton flavor, for the limit computation. We separate the dilepton samples into $\mu\mu$, $\Pe\mu$, and $\Pe\Pe$ subsamples and the trilepton sample into a $\mu\mu\mu$ subsample, an $\Pe\Pe\Pe$ subsample, and a subsample containing all events with mixed lepton flavors.

\begin{table*}[htb]
\begin{center}
\topcaption{Number of events predicted for background processes and observed in the opposite-sign dilepton samples with two or three jets (OS1) and with at least 5 jets (OS2), the same-sign dilepton sample (SS), and the trilepton sample. An entry "--" means that the background source is not applicable to the channel.}\label{tab:yields_ll}

\begin{tabular}{lr@{$\,\pm\,$}lr@{$\,\pm\,$}lr@{$\,\pm\,$}lr@{$\,\pm\,$}l}
\hline
Channel   		&\multicolumn{2}{c}{OS1} & \multicolumn{2}{c}{OS2} & \multicolumn{2}{c}{SS}	& \multicolumn{2}{c}{Trileptons}\\
\hline
$\ttbar$   		&  5.2&1.9	& 80&12	& \multicolumn{2}{c}{--} 	& \multicolumn{2}{c}{--} \\
Single top quark	&  2.5&1.3	& 2.0&1.0	& \multicolumn{2}{c}{--} 	& \multicolumn{2}{c}{--} \\
Z             		&  9.7&2.9 	& 2.5&1.9	& \multicolumn{2}{c}{--} 	& \multicolumn{2}{c}{--} \\
$\ttbar\PW$		&  \multicolumn{2}{c}{--}		& \multicolumn{2}{c}{--} 		&  5.8&1.9		& 0.25&0.11    \\
$\ttbar\Z$ 		&  \multicolumn{2}{c}{--}		& \multicolumn{2}{c}{--} 		&  1.83&0.93  	& 1.84&0.94  \\
\PW\PW  			&  \multicolumn{2}{c}{--}		& \multicolumn{2}{c}{--} 		&  0.53&0.29 	& \multicolumn{2}{c}{--} \\
\PW\Z 			&  \multicolumn{2}{c}{--}		& \multicolumn{2}{c}{--} 		& 0.34&0.08 	& 0.40&0.21  \\
\cPZ\cPZ        		&  \multicolumn{2}{c}{--}		& \multicolumn{2}{c}{--} 		&  0.03&0.00  	& 0.07&0.01  \\
\PW\PW\PW/\PW\PW\cPZ/\cPZ\cPZ\cPZ/\PW\cPZ\cPZ 	&  \multicolumn{2}{c}{--} 		& \multicolumn{2}{c}{--} 		&  0.13&0.07  	& 0.08&0.04  \\
$\ttbar\PW\PW$ 		&  \multicolumn{2}{c}{--}		& \multicolumn{2}{c}{--} 		&  \multicolumn{2}{c}{--}			& 0.05&0.03  \\
Charge misidentification 		&  \multicolumn{2}{c}{--}		& \multicolumn{2}{c}{--} 		&  0.01&0.00 	& \multicolumn{2}{c}{--} \\
Non-prompt 		&  \multicolumn{2}{c}{--}		& \multicolumn{2}{c}{--} 		&  7.9&4.3 		& 0.99&0.90  \\
\hline
Total background	 & 17.4&3.7	&84&12	& 16.5&4.8		& 3.7&1.3 \\
Data               	&  \multicolumn{2}{c}{20}& \multicolumn{2}{c}{86}& \multicolumn{2}{c}{18} & \multicolumn{2}{c}{2} \\
\hline
\end{tabular}
\end{center}
\end{table*}

\begin{table*}[htb]
\begin{center}
\topcaption{Efficiencies $\epsilon$ and number of events $N$ for the T-quark signal with the nominal branching fractions into \bW, \tH, \tZ of 50\%, 25\%, 25\%, respectively, in the opposite-sign dilepton samples with two or three jets (OS1) and with at least 5 jets (OS2), the same-sign dilepton sample (SS), and the trilepton sample.}\label{tab:efficiency_ll}

\begin{tabular}{rcr@{.}lcr@{.}lcr@{.}lcr@{.}l}
\hline
\multicolumn{1}{c}{Channel}          &\multicolumn{3}{c}{OS1}  &\multicolumn{3}{c}{OS2}  &\multicolumn{3}{c}{SS} &\multicolumn{3}{c}{Trileptons} \\
\multicolumn{1}{c}{T mass (\GeVns{})} & $\epsilon$ & \multicolumn{2}{c}{$N$} & $\epsilon$ & \multicolumn{2}{c}{$N$} & $\epsilon$ & \multicolumn{2}{c}{$N$} & $\epsilon$ & \multicolumn{2}{c}{$N$}   \\ \hline

500 & 0.15\% & 16&7 & 0.31\%  &35&1 & 0.19\% &   21&3& 0.17\% &   19&1  \\
600 & 0.27\% & 8&9  & 0.50\%  &16&6 & 0.22\% &   7&5 & 0.26\% &   8&5   \\
700 & 0.36\% & 4&0  & 0.60\%  & 6&6  & 0.25\% &   2&8 & 0.28\% &   3&1   \\
800 & 0.39\% & 1&6  & 0.61\%  & 2&5  & 0.25\% &   1&0 & 0.32\% &   1&3   \\
900 & 0.43\% & 0&67  & 0.60\% & 0&96  & 0.25\% &   0&40 & 0.33\% &	0&52	\\
1000 & 0.44\% & 0&28  & 0.56\% & 0&36  & 0.23\% &   0&15 & 0.33\% &	0&21	\\
1100 & 0.44\% & 0&12  & 0.52\% & 0&14  & 0.22\% &   0&06 & 0.32\% &	0&09	\\
1200 & 0.45\% & 0&05  & 0.46\% & 0&05  & 0.20\% &   0&02 & 0.31\% &	0&04	\\
\hline
\end{tabular}
\end{center}
\end{table*}

\section {Limit computation and systematic uncertainties}
\label{sec:limit}
We observe no evidence for a signal in the data. This section discusses
upper limits on the production cross section of T-quark pairs.
We use Bayesian statistics to compute 95\% confidence level (CL) upper limits for the production cross section for values of the T-quark mass between 500 and 1500\GeV in 100\GeV steps. For the single-lepton channels we compute the posterior probability density as a function of the $\TTbar$ production cross section using the BDT discriminant distribution observed for data at each mass value and the combination of the BDT discriminant distributions for signal and background processes. For the multilepton channels we use the observed and predicted numbers of events in the twelve subsamples to compute the likelihood. We integrate the posterior probability density function over the nuisance parameters assigned to the sources of systematic uncertainties that affect both the normalization and the distribution of the discriminating observables.

Uncertainties in the normalization of the signal and background samples arise from the 2.6\% uncertainty in the integrated luminosity for the $\sqrt{s}=8$\TeV data collected by CMS in 2012~\cite{CMS-PAS-LUM-13-001}, and the uncertainties in the cross sections and in the efficiency corrections. We assign a systematic uncertainty of 50\% for each of  the  diboson backgrounds, for the single-top-quark production, and for the \PW- and \Z-boson backgrounds. This accounts for the uncertainties related to the definition of the renormalization and factorization scales used in the simulation, which is the largest with a systematic uncertainty of 40\%, and for the uncertainties in the determination of the \PW+jets and Drell--Yan backgrounds from data. For the normalization of the $\ttbar$ background we use the NNLO cross section of 245.8\unit{pb}~\cite{Czakon:2013goa} with an 8\% uncertainty to cover the difference between alternative calculations~\cite{Kidonakis:2012db, Cacciari:2011hy}. We correct the lepton trigger and identification efficiencies in the simulation to agree with the performance observed in the data. The uncertainties in the correction factors give rise to uncertainties of 3\% in the normalization of the signal and background samples. We further account for the effect of uncertainties in the jet energy and resolution, the \cPqb-tagging efficiency,  the renormalization and factorization scales, the jet-parton matching scale, and the top-quark-\pt distribution on the number of events expected and the distribution of the BDT discriminant.
The uncertainties related to the PDFs used to model the hard scattering of the proton-proton collisions are determined to be negligible.

\begin{figure}[hbtp]
\begin{center}
\includegraphics[width=0.5\textwidth]{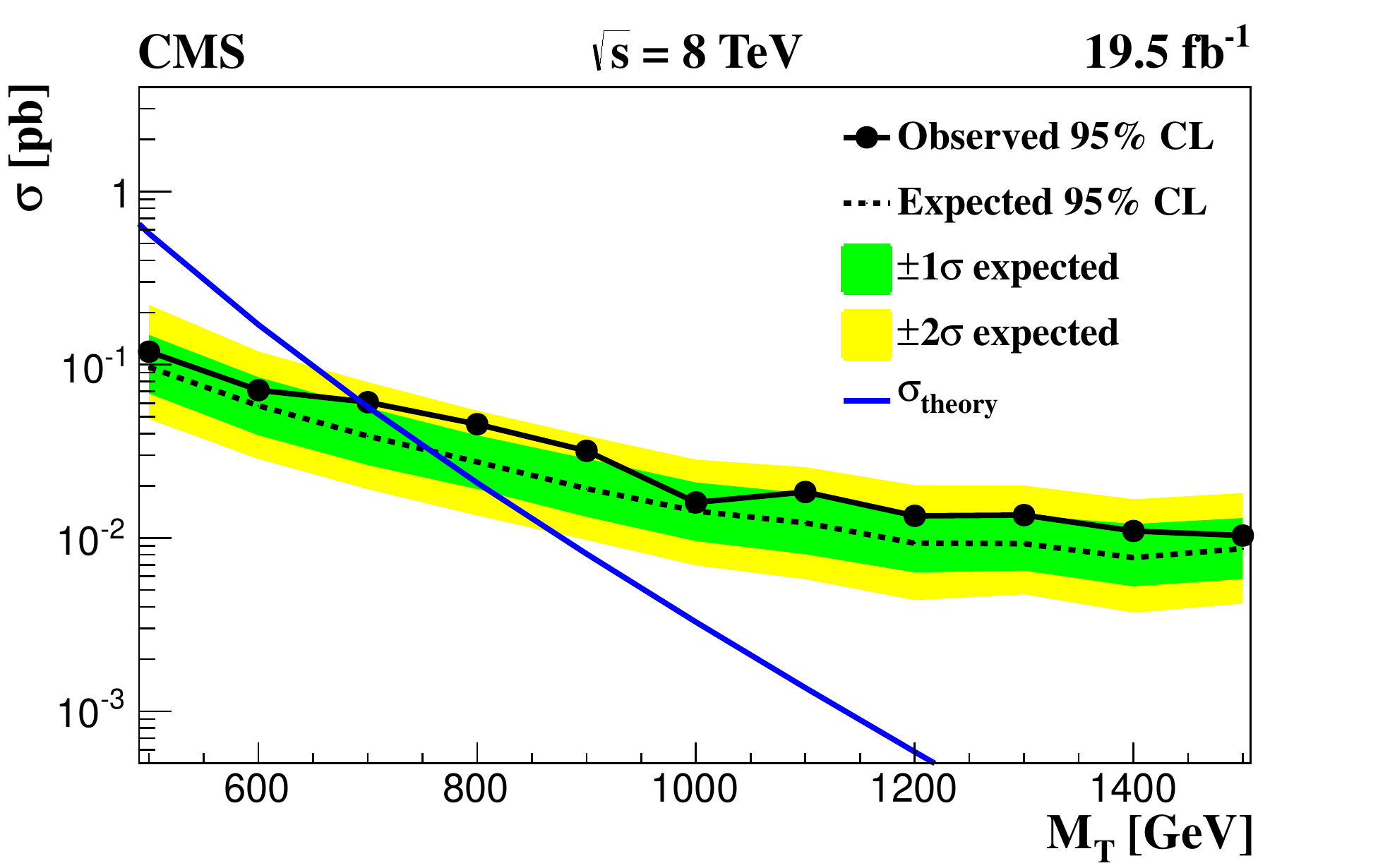}
\caption{Observed and expected 95\% confidence level upper limits for the T-quark production cross section for the nominal branching fractions into \bW, \tH, \tZ of 50\%, 25\%, 25\%, respectively.}
\label{fig:1Dlimit}
\end{center}
\end{figure}

\begin{figure}[hbt]
\begin{center}
\includegraphics[width=\cmsFigWidth]{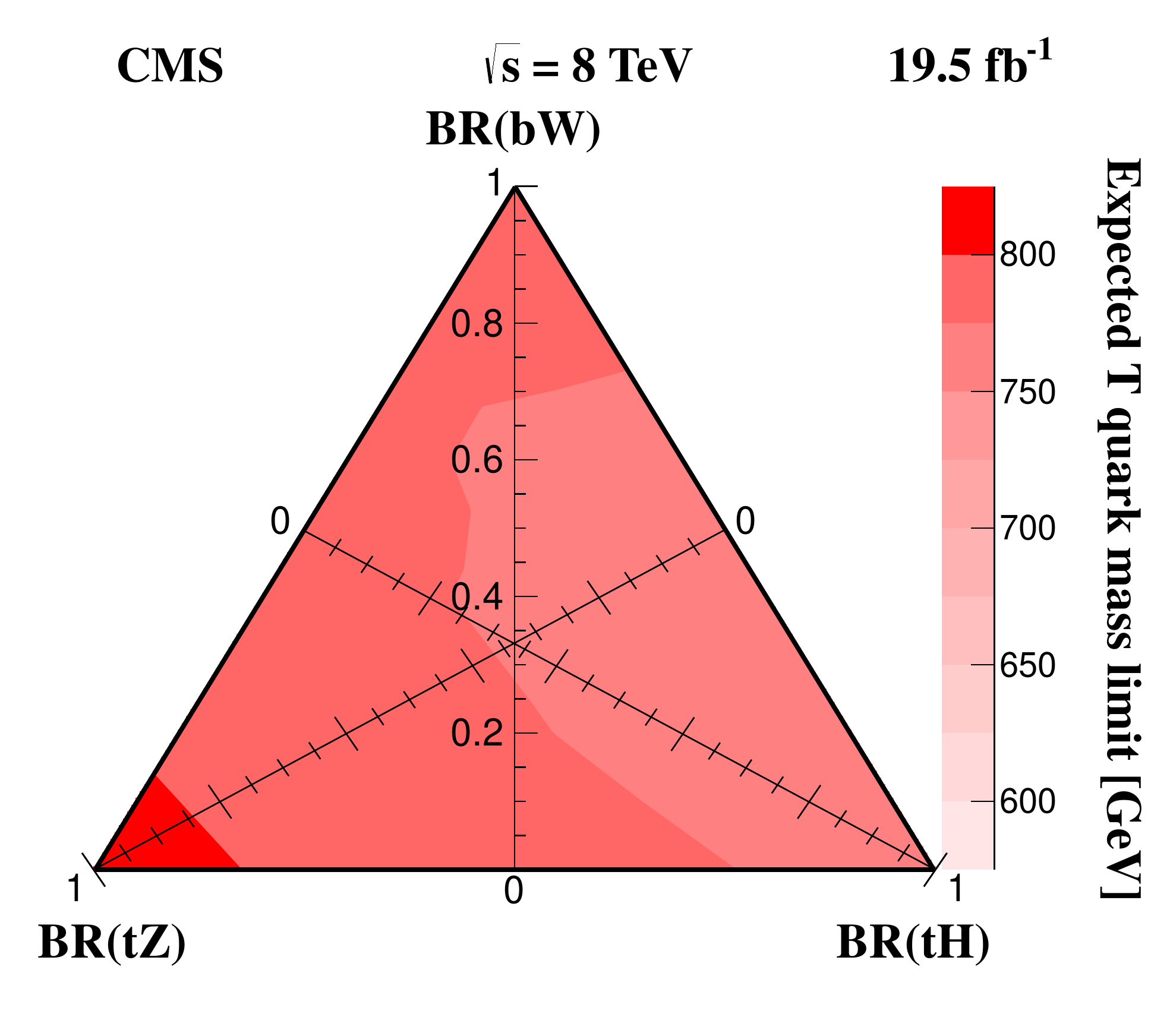}\
\includegraphics[width=\cmsFigWidth]{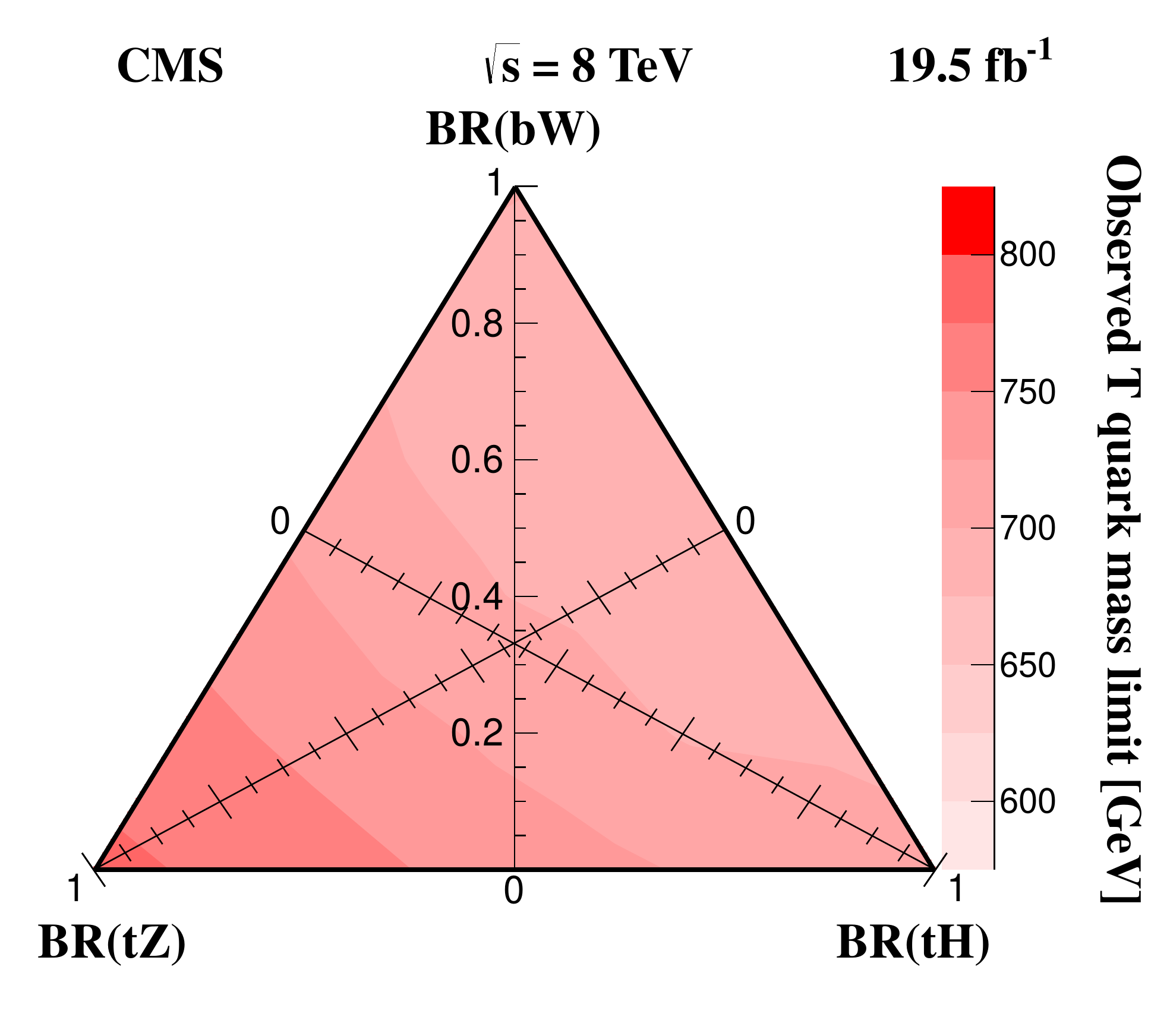}
\caption{Branching-fraction triangle with expected  (\cmsLeft) and observed 95\% CL  limits (\cmsRight) on the T-quark mass. Every point in the triangle corresponds to a specific set of branching-fraction values subject to the constraint that all three add up to 1. The branching fraction for each mode decreases from 1 at the corner labeled with the decay mode to 0 at the opposite side of the triangle.}
\label{fig:triangle}
\end{center}
\end{figure}

The observed and expected limits for the nominal branching fractions are shown in Figure~\ref{fig:1Dlimit}.
The observed limit is slightly higher than expected because there are slightly more events observed than expected in the high tail of the BDT distribution from single-lepton events  with at least one \PW\ jet and in the multilepton channels.
We set a lower limit at the mass of the T quark where the observed cross section limit and the predicted T-quark production cross section intersect.
To model the BDT discriminant distribution expected for different values of the T-quark branching fractions we weight the contributions from the six signal samples according to the branching fractions.
The lower limits for the T-quark mass measured for the different sets of branching fractions are listed in Table~\ref{tab:BR} and represented graphically in Fig.~\ref{fig:triangle}.

\begin{table*}[htb]
\begin{center}
\topcaption{Lower limits for the T quark mass, at 95\% CL, for different combinations of T quark branching fractions.} \label{tab:BR}
\begin{tabular}{ccccc}
\hline
\multicolumn{3}{c}{Branching fractions} & Expected & Observed \\
$\cPQT\rightarrow\bW$ & $\cPQT\rightarrow \tH$ &$\cPQT\rightarrow \tZ$ ~~& limit (\GeVns{}) & limit (\GeVns{}) \\
\hline
0.5 & 0.25 & 0.25 & 773 & 696 \\
0.0 & 0.0 & 1.0 & 813 & 782 \\
0.0 & 0.2 & 0.8 & 798 & 766 \\
0.0 & 0.4 & 0.6 & 790 & 747 \\
0.0 & 0.6 & 0.4 & 783 & 731 \\
0.0 & 0.8 & 0.2 & 773 & 715 \\
0.0 & 1.0 & 0.0 & 770 & 706 \\
0.2 & 0.0 & 0.8 & 794 & 758 \\
0.2 & 0.2 & 0.6 & 786 & 739 \\
0.2 & 0.4 & 0.4 & 777 & 717 \\
0.2 & 0.6 & 0.2 & 767 & 698 \\
0.2 & 0.8 & 0.0 & 766 & 694 \\
0.4 & 0.0 & 0.6 & 786 & 734 \\
0.4 & 0.2 & 0.4 & 776 & 705 \\
0.4 & 0.4 & 0.2 & 766 & 693 \\
0.4 & 0.6 & 0.0 & 762 & 690 \\
0.6 & 0.0 & 0.4 & 779 & 703 \\
0.6 & 0.2 & 0.2 & 771 & 693 \\
0.6 & 0.4 & 0.0 & 769 & 687 \\
0.8 & 0.0 & 0.2 & 779 & 695 \\
0.8 & 0.2 & 0.0 & 777 & 689 \\
1.0 & 0.0 & 0.0 & 785 & 700 \\
\hline
\end{tabular}
\end{center}
\end{table*}

\section {Summary}

We have searched for the associated production of a heavy vector-like T quark with charge $\frac{2}{3}$ and its
antiparticle, based on events with at least one isolated lepton.
No evidence for a signal in the data is seen.
Assuming that the T quark decays exclusively into \bW, \tZ, and \tH, we set lower limits for its mass between 687 and 782\GeV for all possible branching fractions into these three final states assuming strong production. This is the first search that considers all three final states, and these limits are the most stringent constraints to date on the existence of such a quark.

\section*{Acknowledgements}

We congratulate our colleagues in the CERN accelerator departments for the excellent performance of the LHC and thank the technical and administrative staffs at CERN and at other CMS institutes for their contributions to the success of the CMS effort. In addition, we gratefully acknowledge the computing centres and personnel of the Worldwide LHC Computing Grid for delivering so effectively the computing infrastructure essential to our analyses. Finally, we acknowledge the enduring support for the construction and operation of the LHC and the CMS detector provided by the following funding agencies: BMWF and FWF (Austria); FNRS and FWO (Belgium); CNPq, CAPES, FAPERJ, and FAPESP (Brazil); MES (Bulgaria); CERN; CAS, MoST, and NSFC (China); COLCIENCIAS (Colombia); MSES (Croatia); RPF (Cyprus); MoER, SF0690030s09 and ERDF (Estonia); Academy of Finland, MEC, and HIP (Finland); CEA and CNRS/IN2P3 (France); BMBF, DFG, and HGF (Germany); GSRT (Greece); OTKA and NKTH (Hungary); DAE and DST (India); IPM (Iran); SFI (Ireland); INFN (Italy); NRF and WCU (Republic of Korea); LAS (Lithuania); CINVESTAV, CONACYT, SEP, and UASLP-FAI (Mexico); MBIE (New Zealand); PAEC (Pakistan); MSHE and NSC (Poland); FCT (Portugal); JINR (Dubna); MON, RosAtom, RAS and RFBR (Russia); MESTD (Serbia); SEIDI and CPAN (Spain); Swiss Funding Agencies (Switzerland); NSC (Taipei); ThEPCenter, IPST, STAR and NSTDA (Thailand); TUBITAK and TAEK (Turkey); NASU (Ukraine); STFC (United Kingdom); DOE and NSF (USA).
 Individuals have received support from the Marie-Curie programme and the European Research Council and EPLANET (European Union); the Leventis Foundation; the A. P. Sloan Foundation; the Alexander von Humboldt Foundation; the Belgian Federal Science Policy Office; the Fonds pour la Formation \`a la Recherche dans l'Industrie et dans l'Agriculture (FRIA-Belgium); the Agentschap voor Innovatie door Wetenschap en Technologie (IWT-Belgium); the Ministry of Education, Youth and Sports (MEYS) of Czech Republic; the Council of Science and Industrial Research, India; the Compagnia di San Paolo (Torino); the HOMING PLUS programme of Foundation for Polish Science, cofinanced by EU, Regional Development Fund; and the Thalis and Aristeia programmes cofinanced by EU-ESF and the Greek NSRF.

\bibliography{auto_generated}   
\cleardoublepage \appendix\section{The CMS Collaboration \label{app:collab}}\begin{sloppypar}\hyphenpenalty=5000\widowpenalty=500\clubpenalty=5000\input{B2G-12-015-authorlist.tex}\end{sloppypar}
\end{document}

%% file: B2G-12-015-authorlist.tex
\textbf{Yerevan Physics Institute,  Yerevan,  Armenia}\\*[0pt]
S.~Chatrchyan, V.~Khachatryan, A.M.~Sirunyan, A.~Tumasyan
\vskip\cmsinstskip
\textbf{Institut f\"{u}r Hochenergiephysik der OeAW,  Wien,  Austria}\\*[0pt]
W.~Adam, T.~Bergauer, M.~Dragicevic, J.~Er\"{o}, C.~Fabjan\cmsAuthorMark{1}, M.~Friedl, R.~Fr\"{u}hwirth\cmsAuthorMark{1}, V.M.~Ghete, C.~Hartl, N.~H\"{o}rmann, J.~Hrubec, M.~Jeitler\cmsAuthorMark{1}, W.~Kiesenhofer, V.~Kn\"{u}nz, M.~Krammer\cmsAuthorMark{1}, I.~Kr\"{a}tschmer, D.~Liko, I.~Mikulec, D.~Rabady\cmsAuthorMark{2}, B.~Rahbaran, H.~Rohringer, R.~Sch\"{o}fbeck, J.~Strauss, A.~Taurok, W.~Treberer-Treberspurg, W.~Waltenberger, C.-E.~Wulz\cmsAuthorMark{1}
% ...
\vskip\cmsinstskip
% ...
1:~~Also at Vienna University of Technology, Vienna, Austria\\
2:~~Also at CERN, European Organization for Nuclear Research, Geneva, Switzerland\\
% ...
